# Introduction and reconciliation of the ROS and aging paradoxes

Yaguang Ren[1,*], Chao Zhang[1,*]

[1-]Translational Medical Center for Stem Cell Therapy and Institute for Regenerative Medicine, Shanghai East Hospital, Shanghai Key Laboratory of Signaling and Disease Research, School of Life Sciences and Technology, Tongji University, Shanghai, China
*-Correspondence: renyaguang3@163.com (Y. Ren); zhangchao@tongji.edu.cn (C. Zhang)

## Abstract

Researchers in the aging field are getting headache by the ROS and aging paradoxes. There are both experiments supporting and challenging the role of ROS in aging, especially in studies using *C. elegans* as the model. The view that ROS are beneficial and can slow down aging is getting popular. In this paper we proposed explanations on how these contradictive conclusions were generated by taking into account previously unnoticed factors including the excessive response, retardation of growth, and the reliability of ROS measuring approaches. We believe that aging is encoded by genes or DNA and is influenced synthetically by pro-aging factors like ROS and unknown side effects, and anti-aging factors such as particular protective responses. Any pro- or anti-aging roles should be ascribed to the "net" effects rather than that of one particular factor like ROS. From this point of view the ROS and aging paradoxes can be reasonably explained and there are no contradictions with the oxidative stress theory of aging. Nevertheless, we also believe that ROS have limited role in aging considering the prime outcome of evolution or natural selection should be increased adaption to the environment rather than long lifespan. The increase of longevity observed in model organisms may be side effects of retrograde responses motivated against certain circumstances.

## Highlights

- Aging is influenced synthetically by pro-aging factors such as ROS and anti-aging factors such as protective responses.
- The anti-aging effect may be byproduct of retrograde responses motivated against adverse circumstances.
- ROS may be more closely correlated with metabolism rather than aging.
- The prime outcome of evolution should be enhanced adaption to environment rather

than long lifespan.
- ROS levels should be tuned to adjust to metabolism and may play limited role in aging.

**Introduction**

Reactive oxygen species (ROS) are highly reactive chemical species containing oxygen and are mainly generated in mitochondria during respiration (Balaban et al., 2005). ROS react with macromolecules including proteins, lipids, and nucleotides (Balaban et al., 2005; Davalli et al., 2016). The oxidative stress theory of aging speculated that damages caused by ROS would lead to cellular dysfunctions and aging (Harman, 1956). There are experiments both supporting and challenging the theory, and recently the view that ROS are beneficial and can slow down aging is getting more and more popular (Doonan et al., 2008; Hekimi et al., 2011; Lapointe and Hekimi, 2010; Payne and Chinnery, 2015; Scialo et al., 2016). In this paper the ROS and aging paradoxes are briefly summarized and reconciled, which we hope will make things clear in correlated areas.

**Description of the ROS and aging paradoxes**

The "ROS and aging paradoxes" are mainly reflected in the following pairs of contradictive reports:

(a) The antioxidant resveratrol (RSV) slows down aging in both invertebrates and vertebrates (Bhullar and Hubbard, 2015; Hubbard and Sinclair, 2014), and the beneficial effects of other antioxidants including N-acetylcysteine (NAC), vitamin C, reduced glutathione, thioproline, and platinum nanoparticle are also reported (Desjardins et al., 2017; Kim et al., 2008; Shibamura et al., 2009). However, the prooxidant paraquat (PQ) is also considered to have anti-aging effects (Lee et al., 2010; Yang and Hekimi, 2010; Yee et al., 2014).

(b) Over-expression of the major antioxidant enzyme SOD-1 increases and deletion of the thioredoxin TRX-1 decreases longevity in *C. elegans* (Cabreiro et al., 2011; Fierro-Gonzalez et al., 2011; Miranda-Vizuete et al., 2006). Mice with deletion of *sod-1* have accelerated aging phenotype correlated with increased cellular senescence (Zhang et al., 2017), and are ideal models for human frailty (Deepa et al., 2017). But there are also reports showed that deletion all five *sod* genes in *C. elegans* did not decrease lifespan and over-expression of *sod-2* even increased it (Van Raamsdonk and Hekimi, 2009, 2012).

(c) Genetic or environmental perturbations that prolong lifespan including reduced insulin/IGF-1 signaling (IIS), mitochondrial dysfunctions, and dietary restriction (DR) usually activate multiple protective responses such as increased expression of antioxidant and xenobiotic detoxification enzymes, increased autophagy, and other unknown adjustments (Kapahi et al., 2016; Shore et al., 2012; Tullet, 2015). The DAF-16/FoxO3a-dependent longevity signal was also shown to be initiated by antioxidants (Kim et al., 2014). Consistenting with the upregulation of antioxidant enzymes ROS are found decreased in worms chronically treated with sub-lethal levels of the prooxidant paraquat (Lee et al., 2010; Ren et al., 2017; Ren and Zhang, 2017), and in those with deficiencies of genes encoding the mitochondrial subunits NUO-6 or CCO-1 (Yang and Hekimi, 2010). However, some studies reported increased ROS under the above mentioned conditions (Yee et al., 2014).

Researchers are getting puzzled by these paradoxes and are beginning to describe ROS as a beneficial player in aging. We believe that if the pro-longevity roles be attributed to the synthetic effects of secondary responses including the activation of protective mechanism, retardation of growth, and other unknown factors, rather than that of ROS, most if not all of these paradoxes would be reasonably reconciled. Some paradoxes may also be originated from the unreliability of ROS detecting approaches explained in the following words.

## Reconciliation of the ROS and aging paradoxes

The pro-longevity roles should be ascribed to the motivated protective mechanisms rather than ROS

In response to increased ROS levels, protective mechanisms are excessively activated including up-regulated expression of antioxidant enzymes (Ren and Zhang, 2017; Zarse et al., 2012), increased autophagy (Li et al., 2015), and other unknown adjustments. The persistent and excessive activation of these mechanisms may lead to decrease of ROS in the long term as described by the excessive response model (Ren and Zhang, 2017). Among these mechanisms, the antioxidant enzymes and autophagy related pathways are reported to have anti-aging effects (Cabreiro et al., 2011; Fierro-Gonzalez et al., 2011; Miranda-Vizuete et al., 2006). If the protective mechanisms are anti-aging then ROS are likely to be the opposite because the former are motivated as countermeasures against the latter. Just like we consider the immune response but not virus as being beneficial, the pro-longevity effects, if any, should also be ascribed to the protective mechanisms rather than ROS. If so, most of the paradoxes would be reasonable reconciled and the seemingly contradictive studies are actually in accordance with the oxidative theory of aging.

Growth retardation should be taken into account sometimes

Deficiencies of either cytosolic or mitochondrial SODs usually lead to reduced lifespan, sickness, and lethality in yeasts, flies, and mice (Elchuri et al., 2005; Phillips et al., 1989; Unlu and Koc, 2007). But deletion of all *sods* genes in *C. elegans* does not decrease longevity and knockout of the mitochondrial localized superoxide dismutase *sod-2* increases it (Van Raamsdonk and Hekimi, 2009, 2012). Although the *sod* quintuple mutant worms are considered to have normal lifespan they exhibit slow development, reduced fertility, slower defecation cycle, decreased movement, and increased sensitivity to oxidative stresses. The *sod* quintuple mutant and *sod-2* single mutant worms both need more than two days to finish growth compared to wildtype (Van Raamsdonk and Hekimi, 2009, 2012). If growth retardation was taken into account the quintuple mutant worms' lifespan would decrease and the pro-longevity effect of deletion of *sod-2* would diminish. In addition, *sod-2* deletion was reported to have no effect on lifespan in other studies (Doonan et al., 2008). Unlike vertebrates, it is quite common for *C. elegans* to retard growth under detrimental conditions and the retardation is usually accompanied with sickness or fragility (Fraser et al., 2000). Similarly, the *sod* mutants are also sick and vulnerable (Van Raamsdonk and Hekimi, 2009, 2012). Therefore, some of the paradoxes are produced due to the overlook of growth retardation or sickness.

It should be cautious to use ROS data obtained from tissue lysates or isolated organelles

The reliability of ROS detecting approaches is prerequisite for getting any reliable conclusion. Some of the ROS and aging paradoxes indeed arise from opposite ROS results. For example, ROS are found to be decreased in worms with mutations of the mitochondrial respiratory chain subunits *nuo-6* or *isp-1* (Yang and Hekimi, 2010), and in worms treated chronically with un-lethal level of paraquat (Lee et al., 2010; Ren et al., 2017; Ren and Zhang, 2017). But others reported increased ROS by similar treatments (Yee et al., 2014). The contradictions should be caused by big variations of ROS data obtained from isolated mitochondria. In vitro ROS measurements are unreliable due to the following reasons: Firstly, Unlike DNA, RNA, proteins, lipids, and other stable molecules, ROS are unstable, highly reactive, and change rapidly. ROS are mainly generated as byproducts of mitochondrial respiration, and any perturbations of the concentrations of substrates such as pyruvate, oxygen, ADP, and others would dramatically affect the generation or degradation of ROS. Isolated mitochondria are in unphysiological state and the concentrations of substrates for metabolism are all altered (Sanz, 2016). As the byproducts, ROS levels must also change immediately. Secondly, the lysis procedures such as sonication and homogenation would generate heat themselves, disrupt mitochondrial structures, and lead to intracellular or intraorganelle release of iron ions, all of which would generate extra ROS. In addition, $Fe^{2+}$ ions mediated Fenton reaction is an important source of ROS. It is thus recommended to measure ROS in living worms instead of in worm lysates or isolated organelles (Ren et al., 2017). Finally, the so called "Thinking Set" may also influence the reliability of ROS detection. To some researchers, it seems to be logic that prooxidant treatments or disruptions of mitochondria functions would increase ROS. But in fact it is not the case, although transient prooxidant stresses increase ROS, the chronic treatments lead to opposite results due to the excessive response of the antioxidant systems (Ren et al., 2017; Ren and Zhang, 2017). In worms with deficiencies of *nuo-6* or *cco-1* ROS are also reduced (Ren et al., 2017). Consistently, elevated ROS levels due to increased respiration would activate antioxidant enzymes and further decrease ROS in the long term (Zarse et al., 2012). If the ROS data obtained are opposite, it is not surprise that some of the ROS and aging paradoxes are produced.

ROS should be more closely correlated with metabolism and may have limited role in aging

The prime outcome of evolution or natural selection should be enhanced adaption to environment rather than long lifespan. ROS levels should also be tuned to adjust to metabolism. Mitochondria are the main site for ATP production and ROS generation, suggesting the close correlation between metabolism and ROS. Consistently, ROS participate in cell respiration as intermediate products and act as signals in glucose stimulated insulin secretion (Devasagayam et al., 2004; Pi et al., 2007). They are widely distributed in cells and unlikely play distinctive, direct, and prime roles in aging. Aging should be encoded by genes or DNA and ROS may have limited impact on it, which may be one of the reasons for the origin of the paradoxes mentioned above.

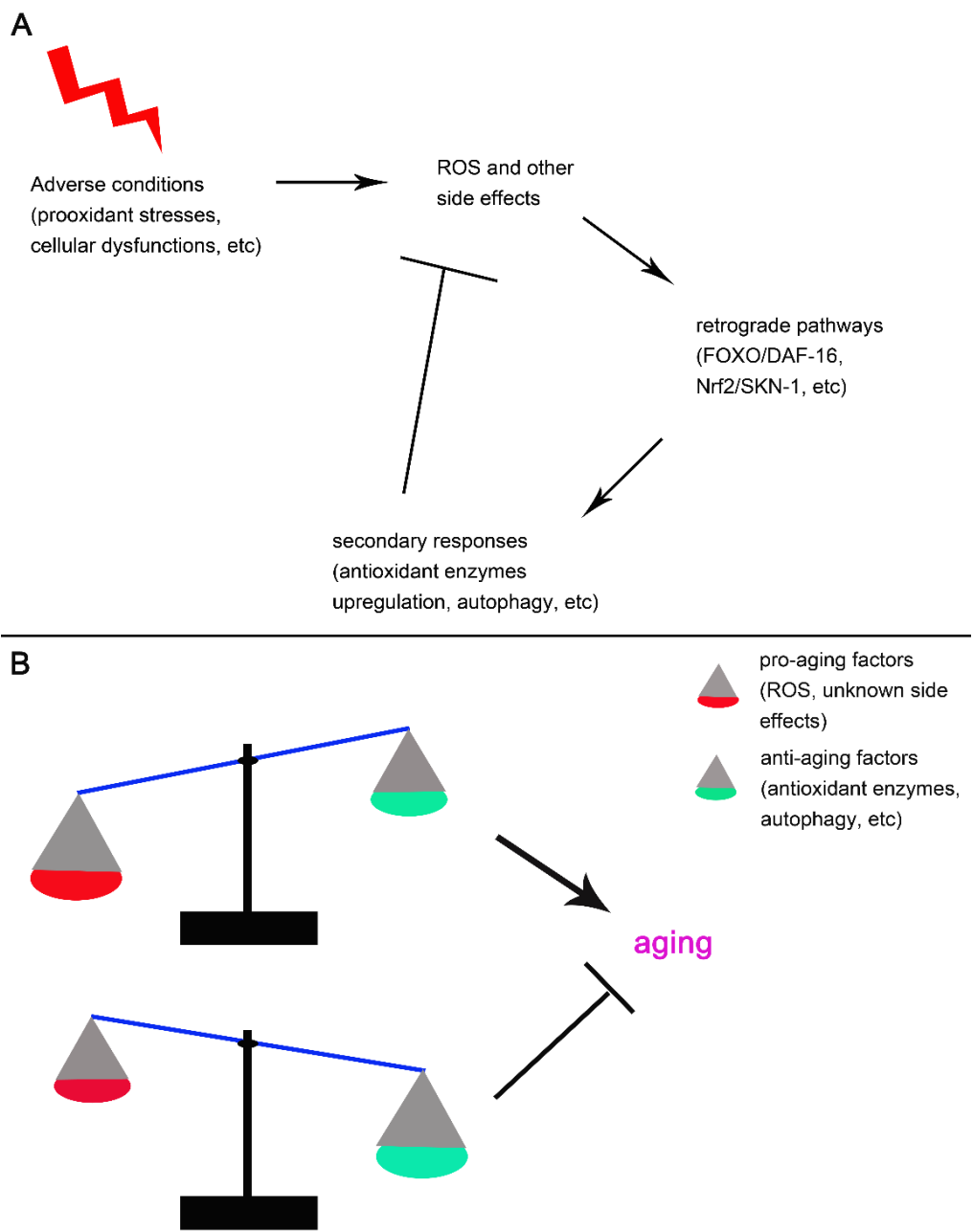

Figure 1. Aging is affected synthetically by pro- and anti-aging factors. (A) Under adverse conditions such as increased prooxidant stresses or cellular dysfunctions ROS and other side effects are produced. Retrograde signaling pathways are activated and downstream secondary responses including upregulation of antioxidant enzymes, increased autophagy, and others are persistently and excessively motivated to fight against ROS or other side effects. (B) If the synthetic effect of pro-aging factors overshadows that of anti-aging factors aging process will be accelerated. Otherwise, the aging process will be decelerated.

## Conclusion

In conclusion, exogenous or endogenous prooxidant stresses would activate adaptive responses including the increased expression of antioxidant enzymes, increased autophagy, and other yet to be identified mechanisms, among which the antioxidant enzymes and autophagy are reported to have anti-aging effects. The excessive response model states that when prooxidant capacity goes high the antioxidant capacity will go higher and lower ROS levels will be observed (Ren and Zhang, 2017). It is thus not surprise that increased longevity is observed under proper level of prooxidant stresses. The protective mechanisms are activated by and tackle against the rise of ROS. If the formers are anti-aging then ROS should be pro-aging, and aging should be affected synthetically by pro-aging factors such as ROS and unknown side effects, and anti-aging factors such as antioxidants, autophagy, and others (Fig.1). Therefore, most if not all of the ROS and aging paradoxes could be reasonably reconciled and we also believe that ROS have limited roles in aging considering that long lifespan is not the prime goal of evolution or natural selection. The increase of longevity observed in model organisms should be byproduct of retrograde responses, of which the main product may be increased adaption to adverse conditions. We are confident that the perspective proposed here will increase the understanding of the relationship between ROS and aging.

## Conflict of interests

The authors declare that there is no conflict of interest.

## Funding

This work was supported by the National Natural Science Foundation of China [Grant numbers 81200253, 81570760, and 31771283]; the National Key Research and Development Program of China [Grant numbers 2017YFA0103900, 2017YFA0103902, and 2016YFA0102200]; One Thousand Youth Talents Program of China to C. Zhang; the Program for Professor of Special Appointment (Eastern Scholar) at Shanghai Institutions of Higher Learning [Grant number A11323]; the Shanghai Rising-Star Program [Grant number 15QA1403600]; and the Fundamental Research Funds for the Central Universities of Tongji University.